\documentclass{pasj00}
\twocolumn

\SetRunningHead{Onodera et al.}{NRO 45-m M\,33 CO survey: II}

\newcommand{\mo}{\: M_\odot}
\newcommand{\pc}{\: \rm pc} 
\newcommand{\kpc}{\: \rm kpc}

\newcommand{\kms}{\: \rm km\,s^{-1}}

\newcommand{\K}{\:\rm K}
\newcommand{\mK}{\:\rm mK}
\newcommand{\cm}{\:\rm cm}

\renewcommand{\micron}{\ensuremath{\mu\mathrm{m}\ }}

\title{NRO M33 All-Disk Survey of Giant Molecular Clouds (NRO MAGiC): II. Dense Gas Formation within Giant Molecular Clouds in M33}
\author{Sachiko~Onodera\altaffilmark{1,2}, Nario~Kuno\altaffilmark{2,3},
Tomoka~Tosaki\altaffilmark{4}, Kazuyuki~Muraoka\altaffilmark{5}, Rie~Miura\altaffilmark{6}, Kotaro~Kohno\altaffilmark{7,8}, Kouichiro~Nakanishi\altaffilmark{3,6},
 Tsuyoshi~Sawada\altaffilmark{9}, Shinya~Komugi\altaffilmark{9}, Hiroyuki~Kaneko\altaffilmark{10}, Akihiko~Hirota\altaffilmark{2}, \\ and \\ Ryohei~Kawabe\altaffilmark{2,9}}

\altaffiltext{1}{Meisei University,  2-1-1 Hodokubo, Hino, Tokyo 191-0042, Japan }
\altaffiltext{2}{Nobeyama Radio Observatory, National Astronomical
Observatory, 462-2 Nobeyama, Minamimaki, Minamisaku, Nagano 384-1305, Japan}
\altaffiltext{3}{The Graduate University for Advanced Studies (Sokendai), 2-21-1 Osawa, Mitaka, Tokyo 181-8588, Japan}
\altaffiltext{4}{Joetsu University of Education, Yamayashiki-machi,
Joetsu, Niigata 943-8512, Japan}
\altaffiltext{5}{Osaka Prefecture University, 1-1 Gakuen-cho, Nakaku, Sakai, Osaka 599-8531, Japan}
\altaffiltext{6}{ALMA Project Office, National Astronomical Observatory, 2-21-1 Osawa, Mitaka, Tokyo 181-8588, Japan}
\altaffiltext{7}{Institute of Astronomy, The University of Tokyo, 2-21-1 Osawa, Mitaka, Tokyo 181-0015, Japan}
\altaffiltext{8}{Research Center for the Early Universe, University of Tokyo, 7-3-1 Hongo, Bunkyo, Tokyo 113-0033, Japan}
\altaffiltext{9}{Joint ALMA Office, Alonso de C\'{o}rdova 3107, Vitacura, Santiago 763-0355, Chile}
\altaffiltext{10}{The University of Tsukuba, 1-1-1 Tennodai, Tsukuba, Ibaraki 305-8577, Japan}

\Received{2011/10/29}
\Accepted{2012/7/2}
\Published{2012/12}

\email{sachiko.onodera@meisei-u.ac.jp}

\maketitle
\begin{document}
\begin{abstract}
We report the results of our observations of the $^{12}$CO\,({\it J}\,=\,1--0) and 
$^{12}$CO\,({\it J}\,=\,3--2) line emission of 74 major giant molecular clouds (GMCs) within the galactocentric distance of $5.1\kpc$ in the Local Group galaxy M33. 
The observations have been conducted as part of the Nobeyama Radio Observatory M33 All-disk 
survey of Giant Molecular
Clouds project (NRO MAGiC). The spatial resolutions are 80pc for $^{12}$CO\,({\it J}\,=\,1--0) and 100pc for 
$^{12}$CO\,({\it J}\,=\,3--2).
We detect $^{12}$CO\,({\it J}\,=\,3--2) emission of 65 GMCs successfully. Furthermore, we find that the correlation 
between the surface density of the star formation rate, which is derived from a linear combination of H$\alpha$ and 24\micron emissions, and the $^{12}$CO\,({\it J}\,=\,3--2)
 integrated intensity still holds at this scale.
This result show that the star-forming activity is closely associated with 
warm and dense gases that are traced with the $^{12}$CO\,({\it J}\,=\,3--2) line, even in the scale of GMCs. We also find that the GMCs with a high star-forming activity tend to show a high integrated intensity ratio ($R_{3-2/1-0}$). 
Moreover, we also observe a mass-dependent trend of $R_{3-2/1-0}$ for the GMCs with a low star-forming activity.
From these results, we speculate that
the $R_{3-2/1-0}$ values of the GMCs with a low star-forming activity mainly depend on the dense gas fraction and not on the temperature, and
therefore, the dense gas fraction increases with the mass of GMCs, at least in the GMCs with a low star-forming activity.
\end{abstract}

\section{Introduction}
The formation processes of massive stars in galaxies are
one of the key topics of study in modern astronomy as these massive stars play an
essential role in the evolution of galaxies. Because stars are formed
from molecular gas, studies on molecular gas in galaxies can 
provide us with crucial information on the fundamental physical
processes of massive star formation. Most of the molecular gas is confined to molecular clouds, and virtually all giant molecular clouds (GMCs) are sites of star formation in the Milky Way. Thus, the physical properties of the gases present in GMCs serve as important information for the investigation of massive star formation.

In order to investigate the evolution of GMCs, it is important to study
GMCs using a statistical approach. For this purpose, observations
of nearby galaxies in the Local Group are preferable, because
we can easily construct a GMC sample at an uniform distance from us 
whereas GMCs in the Milky Way are often subject to the distance determination
issues. In fact, a wide-area $^{12}$CO\,({\it J}\,=\,1--0) survey of the Large Magellanic Cloud (LMC)
provides us with rich information about the variation in GMC properties \citep{fukui08}
with the evolution of the associated star clusters \citep{kawamura09}.

The disk galaxy M33 is another important target for such study because while the LMC is an irregular dwarf galaxy, M33 is one of the nearest galaxies ($D=840\kpc$; \cite{freedman91}) with a spiral morphology. In addition, its favorable inclination angle ($i=52^\circ$; \cite{corbelli00}) enables us to observe the disk face, thereby making this galaxy unique.

Because of these merits, GMCs in M33 have been statistically investigated
through observations of the CO lines in several studies using various existing telescopes
(e.g., for $^{12}$CO\,({\it J}\,=\,1--0): \cite{ws90}; \cite{engargiola03}; \cite{rosolowsky07}, and for $^{12}$CO\,({\it J}\,=\,2--1): \cite{Gratier2011}).
These studies provide us with essential physical properties
of GMCs, including molecular gas masses, sizes, line widths, and
virial parameters. Recent observations using the \textit{Herschel} PACS and SPIRE instruments have 
investigated the far-infrared and sub-millimeter pictures of the cold interstellar medium 
in M33 (HerM33es project: e.g., \cite{kramer2010}).
In addition to these observation results, understanding the formation process of dense molecular gas within GMCs is a critical step in understanding the physical connection between GMC properties and the evolutionary phases of GMCs.
Multi-line observations are indispensable tools for achieving this understanding.
Recent studies have shown that the molecular lines with high critical density 
and/or those of high rotational excitation level are closely associated with massive star formation in external galaxies.  
\citet{kohno99} found that the HCN emission in NGC6951 shows a better spatial correlation  
with the star-forming regions than the $^{12}$CO\,({\it J}\,=\,1--0) emission; \citet{gaosolomon04} 
reported the existence of tight IR-HCN linear correlation in spiral galaxies, which indicates
 that the global star formation rate 
(SFR) is proportional to the mass of dense molecular gas. \citet{mauersberger99} showed the 
relation between $^{12}$CO\,({\it J}\,=\,3--2) emission and the SFR at the centers of 28 nearby galaxies. 
More recently, the 10-m antenna of Atacama Submillimeter-wave Telescope Experiment 
(ASTE: \cite{ezawa04}; \cite{kohno04}) was used to conduct a wide-area survey
of the $^{12}$CO\,({\it J}\,=\,3--2) line toward the inner disk region of the nearby
starburst galaxy M83 \citep{muraoka07},  
revealing an intimate connection between dense molecular gas and
massive star formation at a spatial scale of $\sim\! 500\pc$. 
However, the above mentioned studies are based on observations with resolutions greater than a few $100\pc$, and therefore, observations with higher resolutions are required to study the properties of warm and dense gas in GMCs.

Some case studies of dense gas properties in GMCs have been conducted for nearby
galaxies (\cite{minamidani08}; \cite{Hirota2011}) including M33;
a recent high-resolution ($\sim\!100\pc$) study of NGC604, the most 
active star-forming region in M33, found an the arclike distribution of $^{12}$CO\,({\it J}\,=\,3--2) emission, which is possibly formed through compression by stellar winds and/or 
supernovae shocks of  previously formed stars \citep{tosaki07}. Moreover, aperture-synthesis observations of NGC604 by 
\citet{miura10} revealed that star-forming efficiency decreases as the projected distance 
from the heart of the central cluster increases. Furthermore, they detected HCN emission from two massive 
GMCs at the rim of ``H$\alpha$ shells" produced by the cluster. Their result indicates that 
the evolutionary stages of GMCs change radially in the course of stellar cluster formation. 
Another study of NGC604 using 
a large velocity gradient model for $^{12}$CO\,({\it J}\,=\,1--0), $^{12}$CO\,({\it J}\,=\,3--2), 
and $^{13}$CO\,({\it J}\,=\,1--0) showed that the temperature and density of GMCs 
vary, reflecting their evolutionary stages \citep{Muraoka2012}.

For conducting statistical surveys of dense molecular gas in M33, examination of higher-J CO lines 
such as $^{12}$CO\,({\it J}\,=\,3--2) in the emission spectra of GMCs can be useful, 
because these lines are very bright as compared with the emission lines of high dipole moment molecules
such as HCN and HCO+. For instance, two GMCs in NGC604 whose HCN emissions were detected 
through interferometric observations show a HCN-to-CO flux ratio (flux unit: $\rm Jy\kms$),
$S_{\rm HCN(1-0)}/S_{\rm CO(1-0)}$, of $\sim\! 1/42$ and $1/56$ \citep{miura10}. 
\citet{Rosolowsky2011} reported no detection of HCN emission 
from four GMCs in M33 at $\sim\! 100\pc$ resolution, which set very small upper limits 
of the HCN-to-CO intensity ratio (the intensity is measured in $\K\kms$; hereinafter, the same unit is 
considered for intensity), $I_{\rm HCN(1-0)}/I_{\rm CO(1-0)}$, from $\sim\! 1/40$ down to $1/280$ . 
Furthermore, for other galaxies, \citet{Kuno1995} reported
very small $I_{\rm HCN(1-0)}/I_{\rm CO(1-0)}$ ratios down to $\sim\!1/50$--1/90
in the disk regions of M51, and similar ratios were observed for a few 10-pc scale
observations of GMCs in M31 \citep{Brouillet2005}. 
In contrast, the $I_{\rm CO(3-2)}/I_{\rm CO(1-0)}$ ratios are usually greater than 1/10 even in the disk regions
of galaxies (e.g., \cite{Wielebinski1999}; \cite{Walsh2002}; \cite{muraoka07}); therefore, the $^{12}$CO\,({\it J}\,=\,3--2) line is suitable for whole disk surveys of GMCs.

In this work, we present a study of the warm and dense molecular components of GMCs in M33. 
We investigated these components using a combination of a high-quality On-the-Fly (OTF) 
map of $^{12}$CO\,({\it J}\,=\,3--2) emission obtained using the ASTE 10-m telescope and that of 
$^{12}$CO\,({\it J}\,=\,1--0) emission obtained by the NRO 45-m telescope. 
In this study, we have specifically focused only on the physical properties of GMCs; 
we will use a $^{12}$CO\,({\it J}\,=\,3--2) wide-area image and detailed map-based studies in a
soon-to-be-published paper by \cite{miura12}. This work is part of the Nobeyama Radio Observatory M33 
All-disk Survey of Giant Molecular Clouds (NRO MAGiC) project \citep{tosaki11}. 
 
 \section{Observations and Data Analysis}
\begin{figure*}[h]
\begin{center}
\FigureFile(150mm,200mm){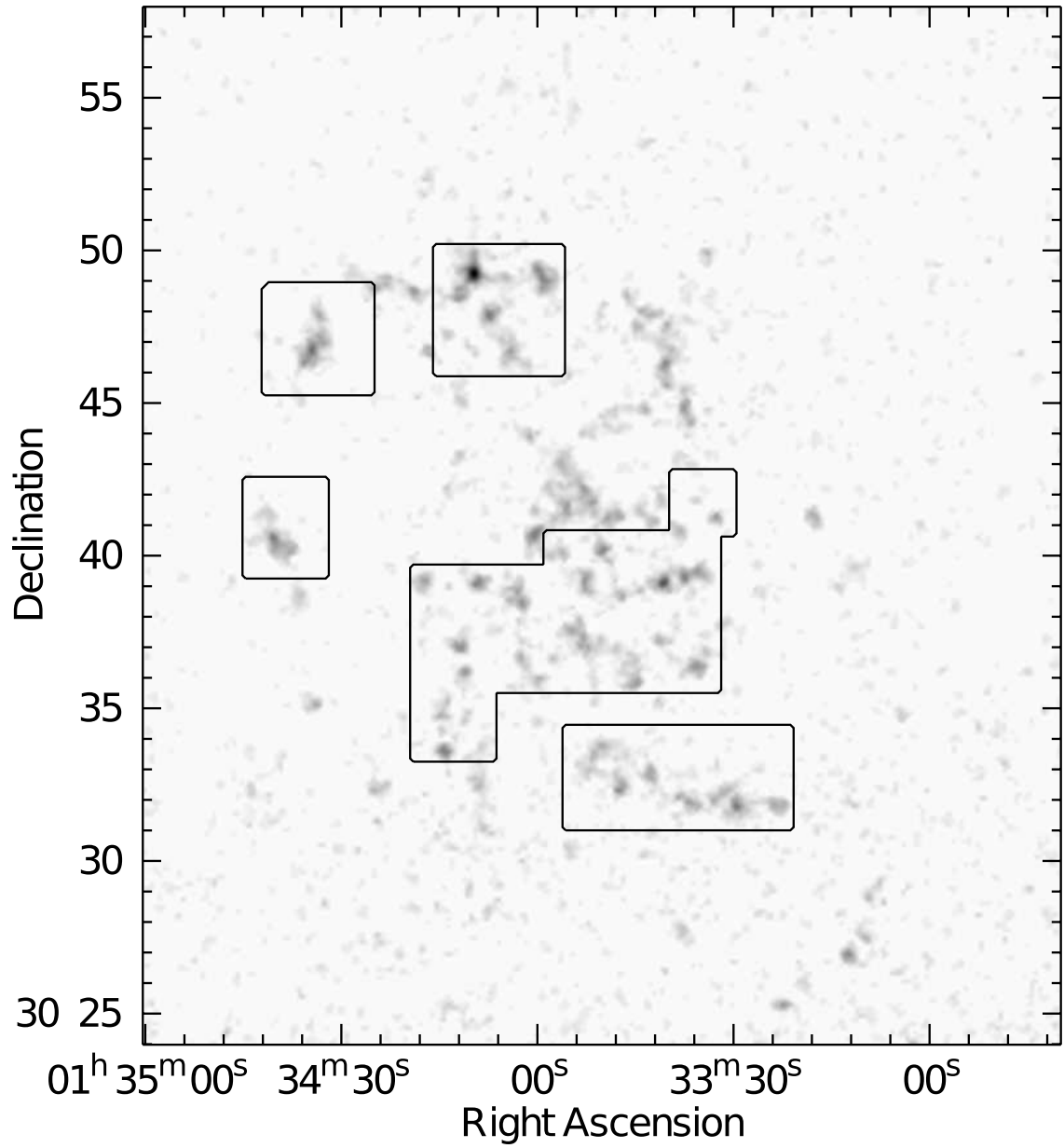}
\end{center}
\caption{Our target regions of the $^{12}$CO\,({\it J}\,=\,3--2) line drawn on the $^{12}$CO\,({\it J}\,=\,1--0) integrated intensity gray-scale map of M33 by \citet{tosaki11}. \label{fig:field}}
\end{figure*}

We first selected target regions that cover the majority of GMCs in M33, as shown in Fig.~\ref{fig:field}.
Subsequently, we conducted $^{12}$CO\,({\it J}\,=\,3--2) observations using the ASTE telescope, 
with a total area coverage of 121 arcmin$^2$ ($\sim\! 7.3\kpc^2$). 
The observations were performed from June to September 2010 and in November 2011 using the OTF observation mode \citep{sawada08}. The front-end component was a CArtridge-Type Sideband-separating receiver for the ASTE 345-GHz band (CATS345: \cite{Inoue2008}). We calibrated the antenna temperature using the standard chopper-wheel technique. The main beam efficiency was $0.6\pm 0.1$ at the frequency of the $^{12}$CO\,({\it J}\,=\,3--2) line.  The temperature scale used throughout this paper is the main-beam brightness temperature ($T_{\rm MB}$).
The typical system noise temperature during the observations was less than $\sim\!200\K$ in 
the single sideband (see \cite{miura12} for a more detailed description of the observations).  

The angular resolution of all images is $25'' (\sim\! 100\pc)$. 
We regridded the images so that their spatial and velocity coordinates matched with the $^{12}$CO\,({\it J}\,=\,1--0) data; the pixel scale is $7''.5$ ($\sim\! 30\pc$) and the velocity resolution is $2.5 \kms$. 
The rms noise was typically 18--$70\mK$. 

We used the $^{12}$CO\,({\it J}\,=\,1--0) data cube from \citet{tosaki11}. This data was obtained 
using the NRO 45-m 
telescope and the 25-beam receiver BEARS. A field of $30'\times 30'$ 
($\sim\! 7.3 \kpc \times 7.3 \kpc$) was mapped using the OTF technique. In that study, the main beam efficiency was $0.32\pm 0.02$ at the frequency of the $^{12}$CO\,({\it J}\,=\,1--0) line.  The errors in the
scaling factors of BEARS were smaller than $\sim\! 20$\%.  
 The data cube had a velocity width of $600\kms$, which adequately covers the full velocity extent of 
the emission in the observed area. Its 
velocity resolution was $2.5\kms$. The angular resolution and the grid spacing were 
$19''.3$ ($\sim\! 80\pc$) and $7''.5$, respectively. The rms noise was $\sigma=130\mK$.

We calculated the SFR using the formulation by \citet{calzetti07}, where the SFR is derived from a linear combination of the H$\alpha$ line and 24-\micron continuum emission. We used the H$\alpha$ image provided by \citet{hoopes00} and the 24-\micron image from the Spitzer Space Telescope data archive. These images were convolved into a $25''$ beam, regridded to a pixel scale of $7''.5$ to match the $^{12}$CO\,({\it J}\,=\,3--2) data, and finally combined.
The rms noises in the H$\alpha$ and 24-\micron maps result in a $\Sigma_{\rm SFR}$ error 
of $4.5\times 10^{-11}{\it \mo}\,\rm \,yr^{-1}\,pc^{-2}$ and a sensitivity limit of $1.4\times 10^{-10}{\it \mo}\,\rm \,yr^{-1}\,pc^{-2}$. 
A more detailed description of the SFR analyses are presented in \citet{onodera10}.

We set the criteria for GMC identification from our $^{12}$CO\,({\it J}\,=\,1--0) data cube as follows: (i) pixels with the $^{12}$CO\,({\it J}\,=\,1--0) brightness temperature $T_{\rm CO(1-0)}$ should be above the noise limit (see below for details)
(ii) all pixels in the cube should be conjunct, (iii) the GMCs should contain at least three pixels at a velocity channel, and (iv) the sizes and velocity widths should be larger than the spatial and velocity resolutions, respectively.
First, we applied the CLUMPFIND algorithm developed by \citet{williams94} to the 
$^{12}$CO\,({\it J}\,=\,1--0) data cube in order to roughly identify the boundaries of the GMCs and 
obtain their masses. The CLUMPFIND algorithm requires a search increment and the lowest 
level that is the same as the noise limit as an input parameter set. 
We searched the most suitable parameter set to minimize misidentification of GMCs and  
assessed the validity of GMC identification by testing many parameter sets and examining 
the resultant channel maps of each GMC candidate. In this process, we rejected 
the parameter sets that resulted in the identification of mere aggregations of widespread small clumps as single large clump candidates. In such cases, the small clumps were not 
connected with each other in either space or velocity domains. We also rejected 
the parameter sets with which most clumps were divided into unreasonably small fragments such as those 
smaller than the spatial and velocity resolutions. 
As a result of this parameter search, we adopted the search increment $1.8\,\sigma$ and the 
lowest level $2.7\,\sigma$ and roughly identified the GMC candidates. 
Subsequently, we removed the isolated pixels that were incorrectly allocated in some of the GMC candidates.
Finally, we rejected the GMC candidates that were close to the rim of the data cube 
as well as the ones whose sizes and velocity widths were smaller than the spatial and velocity resolutions.

We successfully determined 74 GMCs that were also contained in our target fields of $^{12}$CO\,({\it J}\,=\,3--2). Their galactocentric distance ranges from 0.2 to $5.1\kpc$. The H$_2$ gas mass of each GMC is calculated from the $^{12}$CO\,({\it J}\,=\,1--0) data assuming the $^{12}$CO-to-H$_2$ conversion factor $X_{\rm CO}=2\times 10^{20} \cm^{-2}(\K\kms)^{-1}$ \citep{Rosolowsky2003}. The coordinates and velocity at the $T_{\rm CO(1-0)}$ peak position and the H$_2$ gas mass of these GMCs are presented in Table \ref{table:co32param}.

\section{Results}
\subsection{$^{12}$CO\,({\it J}\,=\,1--0) and $^{12}$CO\,({\it J}\,=\,3--2) line spectra\label{co32}}
Fig.~\ref{fig:spectra3210} shows the $^{12}$CO\,({\it J}\,=\,3--2) and $^{12}$CO\,({\it J}\,=\,1--0) spectra at the $T_{\rm CO(1-0)}$ peak positions of each GMC. 
The $^{12}$CO\,({\it J}\,=\,1--0) images were convolved into a $25''$ ($\sim 100\pc$) beam and regridded to a pixel scale of $7''.5$ 
in order to match the $^{12}$CO\,({\it J}\,=\,3--2) data. Our criterion for detection in $^{12}$CO\,({\it J}\,=\,3--2) is that the spectra show emission above the 3-sigma level. The rms noise level for each spectrum is shown in Fig.~\ref{fig:spectra3210}.

Among the 74 GMCs, nine are not detected for $^{12}$CO\,({\it J}\,=\,3--2) emission (Nos. 25, 51, 52, 56, 62, 63, 64, 72, 73). 
Although some of these spectra show multiple peaks, the velocity components of the sub-peaks are excluded from the GMCs in the process of identification. Thus, each GMC corresponds only to the main peak in each spectrum. We fitted the main peaks in these spectra with a single Gaussian to obtain the integrated intensities of CO lines at the peak positions of the GMCs. The integrated intensities are listed in Table \ref{table:co32param}.

\subsection{$^{12}$CO\,({\it J}\,=\,3--2) to $^{12}$CO\,({\it J}\,=\,1--0) integrated intensity ratio $R_{3-2/1-0}$} 
A tight correlation exists between the $^{12}$CO\,({\it J}\,=\,3--2) integrated intensity, $I_{\rm CO\,(3-2)}$, and the $^{12}$CO\,({\it J}\,=\,1--0) integrated intensity, $I_{\rm CO\,(1-0)}$, in nearby galaxies at a $\sim\!500\pc$ resolution \citep{muraoka07}. 
Fig.~\ref{fig:co3210} shows the plot of $I_{\rm CO\,(3-2)}$
versus $I_{\rm CO\,(1-0)}$ of the GMCs in M33. 
We found that the correlation between $I_{\rm CO\,(3-2)}$
and $I_{\rm CO\,(1-0)}$ still holds at the GMC scale with the correlation 
coefficient of 0.65, whereas $R_{3-2/1-0}\equiv I_{\rm CO\,(3-2)}/I_{\rm CO\,(1-0)}$ is 
spread over a wide range from less than 0.1 to 0.74. When weighted by the measurement 
uncertainty, the $R_{3-2/1-0}$ mean is 0.26. The GMCs that are not detected for 
$^{12}$CO\,({\it J}\,=\,3--2) emission are not used for averaging. This value of $R_{3-2/1-0}$ is fairly low as compared with the disk-averaged $^{12}$CO\,({\it J}\,=\,2--1)/$^{12}$CO\,({\it J}\,=\,1--0) of 0.73 discussed by \citet{Gratier2010}. This indicates that the temperature and local volume density at the GMC center is generally not very high at the resolution of $\sim 100\pc$.
Although the GMCs we detected are predominantly located in spiral arms, the $R_{3-2/1-0}$ value is slightly smaller than that of the arm region of the Milky Way galaxy ($\sim\!0.4$, \cite{oka07}).
A GMC with a remarkably high $R_{3-2/1-0}$ value of 0.74 (No.\,13) is located at the center of the most active star forming region, NGC604. 
Such elevated $R_{3-2/1-0}$ values are also reported in the most active HII regions in  LMC, 30\,Dor, and N159 ($R_{3-2/1-0}\simeq\! 0.7$, \cite{minamidani08}), and in larger-scale
measurements of the nearby starburst galaxy M83 
($\sim\! 0.7$, \cite{muraoka07}), luminous IR Galaxies (LIRGs, $\sim\! 0.6$,
\cite{yao03}), and starbursting compact dwarfs ($\sim\! 0.7$, \cite{israel05}). 

\section{Discussion}
In this section, we focus on the physical properties of our sample GMCs and their 
relationship with massive star formation. In particular, we discuss the relationship 
between the denser section of molecular clouds and massive star formation on a scale of
 $\sim\! 100\pc$, as well as the possibility of the existence of a high dense-gas fraction 
within massive GMCs. Here, we used the 
$^{12}$CO\,({\it J}\,=\,1--0) and the SFR data that were convolved into a $\rm 25''(\sim\! 100\pc)$ 
beam so that their resolution was the same as that of the $^{12}$CO\,({\it J}\,=\,3--2) data.

\subsection{Relationship between SFR and CO intensities}
Fig.~\ref{fig:COSFR} shows the plot of $\Sigma_{\rm SFR}$ against $I_{\rm CO\,(1-0)}$ 
and $I_{\rm CO\,(3-2)}$ at the $T_{\rm CO(1-0)}$ peak position in each GMC. 
It should be noted that $\Sigma_{\rm SFR}$ is integrated over all the velocity components in the line of sight, whereas the $I_{\rm CO\,(1-0)}$ and $I_{\rm CO\,(3-2)}$ values correspond only to the velocity extent of the GMC. Because the sub-components in the CO spectra of most GMCs are smaller than the main component, this effect on $\Sigma_{\rm SFR}$ is estimated to be a factor of less than 2.

This figure shows a clear correlation between $I_{\rm CO\,(3-2)}$
and $\Sigma_{\rm SFR}$, with the correlation coefficient being 0.68. In contrast, 
the correlation between $I_{\rm CO\,(1-0)}$ and $\Sigma_{\rm SFR}$ is not obvious, with the correlation coefficient being 0.22. 
This result is consistent with that obtained by \citet{yao03}; their study showed that 
the far-infrared (FIR) luminosity and the $^{12}$CO\,({\it J}\,=\,3--2) luminosity
were better correlated than the FIR luminosity and the $^{12}$CO\,({\it J}\,=\,1--0) luminosity 
in infrared-luminous galaxies with resolutions of $\sim\! 1-14\kpc$. 
Our result is also consistent with that of 
\citet{komugi07}; they presented an identical result for the nuclei of dozens of nearby
galaxies, with a typical resolution of $\sim\! 3\kpc$, and obtained a correlation in
the $I_{\rm CO(3-2)}$ range of $\sim\! 10-10^3\K\kms$. We found
that this correlation is valid down to the $I_{\rm CO(3-2)}$ range of $\sim\! 1-10\K\kms$, 
as shown in Fig.~\ref{fig:COSFR},
even in the GMC scale of $\sim\, 100\pc$. 

This correlation can be attributed to the higher density and/or temperature of the gas 
that is traced with the $^{12}$CO\,({\it J}\,=\,3--2) line.
That is, because the $^{12}$CO\,({\it J}\,=\,3--2) line traces denser gas 
(with a critical density of $n_{\rm crit}\sim 10^4 \rm cm^{-3}$) than 
that traced by the $^{12}$CO\,({\it J}\,=\,1--0) line, with $n_{\rm crit}\sim 10 \rm cm^{-3}$,
a good correlation is expected, as seen between the HCN intensity and the SFR, 
even up to the dense core scale \citep{wu05}. Furthermore, star-forming activities heat 
the surrounding gas, and consequently, the CO molecules in the gas are excited to show 
stronger {\it J}\,=\,3--2 line emission. Because the CO $J=3$ level is 33 K above the ground state 
and the CO $J=1$ level is 5.5 K, we can also expect that the correlation between the SFR and 
the $^{12}$CO\,({\it J}\,=\,3--2) line emission is better than that between the SFR 
and $^{12}$CO\,({\it J}\,=\,1--0) line emission.

As we have shown in the previous subsection, the integrated intensities of $^{12}$CO\,({\it J}\,=\,1--0) and 
$^{12}$CO\,({\it J}\,=\,3--2) are well correlated with each other, at the peak positions of $T_{\rm CO(1-0)}$ in each GMC, 
at a resolution of $\sim 100$pc. Further, considering that $\Sigma_{\rm SFR}$ and $I_{\rm CO\,(3-2)}$ are
 better correlated than $\Sigma_{\rm SFR}$ and $I_{\rm CO\,(1-0)}$, we expect a radial 
gradient of emissions of the CO lines and the SFR tracers in the GMCs. From the center of the GMCs to that of the 
HII regions, we observe peaks of $^{12}$CO\,({\it J}\,=\,1--0), $^{12}$CO\,({\it J}\,=\,3--2), and 
H$\alpha$+24$\mu$m. The existence of the radial gradient can be inspected by directly comparing the maps of $^{12}$CO\,({\it J}\,=\,1--0), $^{12}$CO\,({\it J}\,=\,3--2), and the SFR tracers. The results of this comparison will be presented in our forthcoming paper.

\subsection{Relationship of $R_{3-2/1-0}$ with GMC mass and SFR}
As mentioned above, the emission of the $^{12}$CO\,({\it J}\,=\,3--2) line is more closely related to the SFR than that of the $^{12}$CO\,({\it J}\,=\,1--0) line is. Moreover, it is probable that $R_{3-2/1-0}$ depends on the properties of the GMCs, such as their mass.  
For example, in our galaxy, clouds with masses greater than $10^3\mo$
have denser and warmer cores than those with masses less than $10^3\mo$ (\cite{vk88}).  

Fig.~\ref{fig:R3210-Mgas} shows a plot of the H$_2$ gas mass of the GMCs against 
$R_{3-2/1-0}$ towards the $T_{\rm CO(1-0)}$ peak position of each GMC. 
In this figure, the GMCs are denoted by circles, whose area is
 proportional to the $\Sigma_{\rm SFR}$ at the same position. The red circles
indicate the GMCs with a high star-forming activity ($\Sigma_{\rm SFR} >
1\times 10^{-8}\mo\pc^{-2}$), and the blue ones indicate those with a low
star-forming activity ($\Sigma_{\rm SFR} < 1\times
10^{-8}\mo\pc^{-2}$). Here, in order to check whether $R_{3-2/1-0}$ shows mass-dependency, 
we divide the sample in half at $M_{\rm H2}=8.4\times 10^4\mo$. The crosses indicate the averages of $R_{3-2/1-0}$ in each mass bin; these averages are taken for GMCs with high and low star-forming activities. 
The averages of $R_{3-2/1-0}$ are 0.31 for the 14 less-massive and star-forming GMCs, 0.37 for the 23 massive and star-forming GMCs, and 0.24 for the 14 massive and low star-forming GMCs. 
It should be noted that $R_{3-2/1-0}=0.14$ for the 23 less star-forming, 
less-massive GMCs is a value averaged over upper-limit values.

We found that the GMCs with high star-forming activity tend to show a high $R_{3-2/1-0}$ value. 
Moreover, the mass-dependent trend of $R_{3-2/1-0}$ in the GMCs with a low star-forming activity  is apparent. 
We found that $\sim 24$\% of the GMCs with $M_{\rm H2} < 8.4\times 10^4\mo$ show $R_{3-2/1-0}\leq 0.1$, whereas none of 
the GMCs with $M_{\rm H2} > 8.4\times 10^4\mo$ have such a small value of $R_{3-2/1-0}$. 
This difference between the massive and less-massive GMCs is the reason for the appearance of a blank in the lower-right portion of 
this plot. It also appears in the averaged $R_{3-2/1-0}$: the difference between the mean upper-limit value of 0.14 for less star-forming, 
less-massive GMCs and the mean value of 0.24 for less star-forming, massive GMCs is significant: it is larger than the measurement uncertainties. 
Because the $\Sigma_{\rm SFR}$ in the less star-forming GMCs does not greatly depend on GMC mass,
the mass-dependent trend of $R_{3-2/1-0}$ is attributed to factors other than  $\Sigma_{\rm SFR}$.
On the other hand, $R_{3-2/1-0}$ in the GMCs with a high star-forming activity does not significantly differ 
between the massive and less-massive GMCs: the difference between the average values of $R_{3-2/1-0}$, i.e., 0.37 and 0.31, is comparable with the measurement uncertainties.

On the basis of these results, we speculate that
the values of $R_{3-2/1-0}$ for the GMCs with a low star-forming activity mainly depend on the dense gas fraction and not on the temperature, 
because they have a relatively small amount of energy input from the heating source. 
Considering the mass-dependent trend of the $R_{3-2/1-0}$ values of the GMCs with a low star-forming activity, 
we can state that the dense gas fraction increases with the GMC mass, at least in the case of GMCs with a low star-forming activity.
Because the SFR and dense gas mass show a linear correlation \citep{wu05},
the mass-dependency of the dense gas fraction is consistent with the results obtained by \citet{kennicutt07}, who showed that the star-formation efficiency increases (or the star formation timescale reduces) with an increase in the cloud mass.
On the other hand, $R_{3-2/1-0}$ of the GMCs with star-forming activity does not significantly differ between the massive and less-massive GMCs, because it depends mainly on the temperature.

\section{Summary and Conclusions} 

We identified 74 GMCs from our $^{12}$CO\,({\it J}\,=\,1--0) data of M33. Among these, 
$^{12}$CO\,({\it J}\,=\,3--2) emission of 65 GMCs are detected. 
We compared the $^{12}$CO\,({\it J}\,=\,1--0) and $^{12}$CO\,({\it J}\,=\,3--2) line 
emissions of the GMCs to investigate the variation in their physical properties and 
evolutionary stages. 

\begin{enumerate}

\item The $^{12}$CO\,({\it J}\,=\,1--0) and $^{12}$CO\,({\it J}\,=\,3--2) integrated intensity ratio
      $R_{3-2/1-0}$ is spread over a wide range from less than 0.1 to 0.74, having a weighted mean of 
$R_{3-2/1-0}=0.26$. This weighted mean is slightly smaller than that of the quiescent disk region of the 
Milky Way. 

\item The surface density of SFR and the $^{12}$CO\,({\it J}\,=\,3--2) integrated intensity at the $T_{\rm CO(1-0)}$ peak 
position of GMCs show a good correlation, which is consistent with the 
      findings of previous studies that used data of lower resolutions ($\sim\! 500\pc$).
      We have shown that the correlation is valid down
      to $I_{\rm CO\,({\it J}\,=\,3-2)}\sim\!1$--$10\K\kms$, even in the scale of GMCs. 

\item We have investigated the relationship among the GMC mass,
      $R_{3-2/1-0}$, and the SFR. 
      We have found that the GMCs with a high star-forming activity tend to show a high $R_{3-2/1-0}$ 
value. We have also observed the mass-dependent trend of $R_{3-2/1-0}$ in the GMCs with a low 
star-forming activity. 
On the basis of these results, we speculate that
the $R_{3-2/1-0}$ values of the GMCs with a low star-forming activity mainly depend on the dense gas fraction and not on the temperature. 
This implies that the dense gas fraction increases with the mass of GMCs, at least in the GMCs with a low star-forming activity.

\end{enumerate}
We would like to thank the anonymous referee for his/her comments, which have helped enhance this paper. The Nobeyama Radio Observatory is a branch
of the National Astronomical Observatory of Japan, National Institutes of Natural Sciences.

\begin{longtable}{rcccccccc}
\caption{Observed properties of GMCs \label{table:co32param}}
\hline
\hline
No. & RA & Decl & Vrad & $I_{\rm CO({\it J}\,=\,1-0)}$ & $I_{\rm CO\,({\it J}\,=\,3-2)}$  & $R_{31}$  
& GMC mass $M_{\rm H_2}$& $\rm 10^9\times\Sigma_{SFR}$ \\
&(J2000)&(J2000)&($\rm km\,s^{-1}$)& (K$\kms$) & (K$\kms$)  &   & ($10^4\mo$) & ($\mo \rm yr^{-1}pc^{-2}$)  \\
(1) & (2) & (3) & (4) & (5) & (6) & (7) & (8) & (9) \\
\hline
\endhead
\hline
\endfoot
\hline
\multicolumn{3}{l}{\hbox to 0pt{\parbox{180mm}{\footnotesize  NOTES.---Col. (1): Running Number. 
Col.(2), (3) and (4): The peak position of $T_{\rm CO(1-0)}$ in GMCs. 
Col.(5):The $^{12}$CO\,({\it J}\,=\,1--0) integrated intensity on a $T_{\rm MB}$ scale 
derived by using a single Gaussian fitting for a spectrum at the $T_{\rm CO(1-0)}$ 
peak position in GMCs. The data were convolved into a $\rm 25''(\sim\! 100\pc)$ beam to match
the resolution of the $^{12}$CO\,({\it J}\,=\,3--2) data. Col.(6) Same as col.(5), but for
$^{12}$CO\,({\it J}\,=\,3--2) line. Col.(7): Ratio of $I_{\rm CO\,(3-2)}$ to 
$I_{\rm CO\,(1-0)}$. Col.(8): The H$_2$ gas mass of GMCs. 
Col.(9): 
\footnotemark[$*$] For GMCs whose $^{12}$CO\,({\it J}\,=\,3--2) emissions are not detected, $I_{\rm CO({\it J}\,=\,3-2)}$ is shown as the upper limit, which is estimated as three times the rms of $I_{\rm CO({\it J}\,=\,3-2)}$ that is calculated from the rms noise of $T_{\rm CO({\it J}\,=\,3-2)}$ and the width of $^{12}$CO\,({\it J}\,=\,1--0) line. }}}
\endlastfoot
  1 & 1 34 09.65 & 30 49 07.5 & -248.8 & 21.53 & 5.66 & 0.26 $\pm$ 0.02 & 66.3 & 	  3.2\\
  2 & 1 34 13.66 & 30 33 45.0 & -157.2 & 12.33 & 6.41 & 0.52 $\pm$ 0.07 & 20.9 & 	 36.9\\
  3 & 1 34 06.73 & 30 47 45.0 & -256.7 & 11.34 & 3.14 & 0.28 $\pm$ 0.03 & 24.1 & 	 22.5\\
  4 & 1 33 35.90 & 30 39 30.0 & -167.9 &  9.36 & 4.36 & 0.47 $\pm$ 0.05 & 38.5 & 	 27.9\\
  5 & 1 33 49.85 & 30 33 52.5 & -133.7 &  7.46 & 2.64 & 0.35 $\pm$ 0.04 & 16.2 & 	 13.3\\
  6 & 1 34 10.77 & 30 36 15.0 & -159.3 &  7.80 & 3.41 & 0.44 $\pm$ 0.03 & 13.8 & 	 20.8\\
  7 & 1 33 59.75 & 30 49 15.0 & -252.0 & 10.31 & 3.75 & 0.36 $\pm$ 0.03 & 17.3 & 	  6.0\\
  8 & 1 33 35.91 & 30 36 30.0 & -134.3 &  9.81 & 3.30 & 0.34 $\pm$ 0.04 & 26.3 & 	 20.6\\
  9 & 1 33 41.14 & 30 39 15.0 & -165.8 & 14.89 & 4.68 & 0.31 $\pm$ 0.03 & 35.6 & 	  8.9\\
 10 & 1 34 34.66 & 30 46 15.0 & -221.2 & 11.70 & 2.84 & 0.24 $\pm$ 0.03 & 14.2 & 	 17.4\\
 11 & 1 33 58.58 & 30 48 52.5 & -245.9 & 11.98 & 4.18 & 0.35 $\pm$ 0.04 & 33.3 & 	 11.5\\
 12 & 1 34 08.45 & 30 39 15.0 & -195.0 & 10.75 & 3.24 & 0.30 $\pm$ 0.03 & 21.4 & 	 18.1\\
 13 & 1 34 33.50 & 30 46 45.0 & -241.7 & 13.68 & 10.11 & 0.74 $\pm$ 0.07 & 40.9 & 	263.9\\
 14 & 1 33 29.54 & 30 32 00.0 & -133.2 & 11.49 & 4.41 & 0.38 $\pm$ 0.05 & 23.2 & 	 27.2\\
 15 & 1 34 16.59 & 30 39 15.0 & -190.1 &  8.84 & 1.67 & 0.19 $\pm$ 0.03 & 18.9 & 	  6.2\\
 16 & 1 34 39.27 & 30 40 37.5 & -202.5 & 11.60 & 2.47 & 0.21 $\pm$ 0.02 & 17.8 & 	  2.2\\
 17 & 1 33 44.63 & 30 36 00.0 & -139.3 &  8.17 & 3.85 & 0.47 $\pm$ 0.06 & 18.2 & 	 17.3\\
 18 & 1 34 02.64 & 30 39 00.0 & -199.1 &  8.61 & 2.48 & 0.29 $\pm$ 0.04 &  8.1 & 	 24.0\\
 19 & 1 34 12.51 & 30 37 07.5 & -174.8 & 14.76 & 0.96 & 0.06 $\pm$ 0.02 &  4.3 & 	  5.8\\
 20 & 1 33 43.47 & 30 33 00.0 & -119.2 &  7.72 & 1.45 & 0.19 $\pm$ 0.03 & 15.5 & 	 12.7\\
 21 & 1 34 32.34 & 30 47 45.0 & -243.7 &  8.29 & 3.64 & 0.44 $\pm$ 0.06 & 10.1 & 	 31.3\\
 22 & 1 33 52.76 & 30 39 15.0 & -168.1 &  8.51 & 5.01 & 0.59 $\pm$ 0.09 & 11.4 & 	 48.3\\
 23 & 1 33 47.53 & 30 32 37.5 & -118.1 &  9.97 & 2.69 & 0.27 $\pm$ 0.03 & 15.9 & 	 10.3\\
 24 & 1 33 52.76 & 30 37 30.0 & -154.8 &  7.11 & 2.33 & 0.33 $\pm$ 0.04 & 13.5 & 	  6.4\\
 25 & 1 34 14.25 & 30 34 22.5 & -175.7 &  6.80 & $<1.47$ & $<0.22$ &  3.9 & 	 11.8\\
 26 & 1 33 38.81 & 30 41 00.0 & -186.9 &  6.67 & 1.10 & 0.16 $\pm$ 0.05 &  8.3 & 	  7.6\\
 27 & 1 34 02.06 & 30 38 37.5 & -185.2 & 10.32 & 4.23 & 0.41 $\pm$ 0.05 & 12.6 & 	 43.0\\
 28 & 1 34 38.68 & 30 40 15.0 & -199.3 &  9.28 & 1.14 & 0.12 $\pm$ 0.02 & 20.4 & 	  1.1\\
 29 & 1 33 37.08 & 30 32 07.5 & -120.2 & 10.55 & 2.99 & 0.28 $\pm$ 0.04 & 19.0 & 	 10.0\\
 30 & 1 34 04.41 & 30 49 00.0 & -252.0 &  6.08 & 1.58 & 0.26 $\pm$ 0.05 &  5.4 & 	  2.2\\
 31 & 1 33 50.43 & 30 37 22.5 & -149.0 &  5.56 & 2.29 & 0.41 $\pm$ 0.07 &  6.7 & 	 15.0\\
 32 & 1 33 33.57 & 30 41 15.0 & -185.0 &  8.17 & 3.16 & 0.39 $\pm$ 0.05 &  9.0 & 	102.9\\
 33 & 1 33 34.75 & 30 37 30.0 & -140.5 &  5.51 & 0.59 & 0.11 $\pm$ 0.04 &  2.5 & 	  9.4\\
 34 & 1 34 10.81 & 30 48 22.5 & -252.9 &  9.27 & 1.31 & 0.14 $\pm$ 0.02 & 18.0 & 	  3.5\\
 35 & 1 33 49.85 & 30 40 15.0 & -197.0 & 12.33 & 2.64 & 0.21 $\pm$ 0.03 &  9.4 & 	 10.3\\
 36 & 1 34 11.35 & 30 37 00.0 & -169.0 &  9.80 & 1.68 & 0.17 $\pm$ 0.03 & 11.1 & 	  4.7\\
 37 & 1 33 59.15 & 30 36 07.5 & -154.5 &  7.45 & 3.07 & 0.41 $\pm$ 0.09 & 12.8 & 	 23.7\\
 38 & 1 33 31.28 & 30 31 45.0 & -132.8 &  6.97 & 1.55 & 0.22 $\pm$ 0.04 & 14.6 & 	 18.9\\
 39 & 1 34 13.13 & 30 47 07.5 & -243.8 &  5.51 & 0.97 & 0.18 $\pm$ 0.04 &  5.0 & 	  3.6\\
 40 & 1 34 03.80 & 30 38 37.5 & -187.2 &  6.97 & 2.07 & 0.30 $\pm$ 0.06 &  7.1 & 	 16.3\\
 41 & 1 33 24.31 & 30 32 00.0 & -123.2 &  8.21 & 2.26 & 0.28 $\pm$ 0.04 & 20.4 & 	  3.4\\
 42 & 1 33 54.50 & 30 37 37.5 & -166.0 &  8.53 & 2.38 & 0.28 $\pm$ 0.03 & 11.8 & 	  7.2\\
 43 & 1 33 47.53 & 30 38 52.5 & -157.3 &  6.78 & 1.83 & 0.27 $\pm$ 0.04 &  9.5 & 	 20.0\\
 44 & 1 34 03.82 & 30 46 37.5 & -242.9 &  9.92 & 1.78 & 0.18 $\pm$ 0.02 & 12.3 & 	  5.2\\
 45 & 1 33 34.75 & 30 36 52.5 & -136.7 &  5.70 & 1.91 & 0.34 $\pm$ 0.06 &  4.6 & 	 16.3\\
 46 & 1 33 48.11 & 30 37 07.5 & -145.4 &  5.04 & 0.77 & 0.15 $\pm$ 0.04 &  3.2 & 	  7.9\\
 47 & 1 34 02.05 & 30 36 45.0 & -154.7 &  8.07 & 1.69 & 0.21 $\pm$ 0.03 & 10.7 & 	 10.3\\
 48 & 1 33 35.90 & 30 41 45.0 & -191.5 &  5.05 & 1.09 & 0.22 $\pm$ 0.06 &  4.1 & 	 31.0\\
 49 & 1 33 52.76 & 30 38 52.5 & -155.0 & 10.72 & 3.68 & 0.34 $\pm$ 0.05 &  5.3 & 	 28.8\\
 50 & 1 33 57.41 & 30 36 45.0 & -166.4 &  5.86 & 0.99 & 0.17 $\pm$ 0.04 &  4.9 & 	  3.5\\
 51 & 1 33 53.34 & 30 32 45.0 & -93.3 &  2.85 & $<0.18$ & $<0.06$ &  2.8 & 	  8.3\\
 52 & 1 34 10.77 & 30 37 45.0 & -166.3 &  3.62 & $<0.22$ & $<0.06$ &  2.9 & 	  2.8\\
 53 & 1 34 06.15 & 30 49 37.5 & -255.6 &  6.75 & 0.71 & 0.10 $\pm$ 0.02 &  4.3 & 	  1.6\\
 54 & 1 33 56.25 & 30 40 07.5 & -194.2 &  6.59 & 1.85 & 0.28 $\pm$ 0.04 &  8.4 & 	  6.0\\
 55 & 1 33 51.02 & 30 33 45.0 & -131.4 &  9.76 & 1.96 & 0.20 $\pm$ 0.04 &  6.5 & 	 12.1\\
 56 & 1 33 38.25 & 30 31 22.5 & -103.8 &  4.59 & $<0.16$ & $<0.03$ &  1.9 & 	  6.6\\
 57 & 1 34 10.81 & 30 49 45.0 & -260.8 &  5.39 & 0.79 & 0.15 $\pm$ 0.04 &  7.6 & 	  1.3\\
 58 & 1 34 15.46 & 30 46 37.5 & -235.4 &  6.37 & 0.61 & 0.10 $\pm$ 0.03 &  6.3 & 	  2.7\\
 59 & 1 33 41.14 & 30 37 15.0 & -144.0 &  7.03 & 1.64 & 0.23 $\pm$ 0.05 &  3.4 & 	  7.2\\
 60 & 1 34 13.67 & 30 35 00.0 & -170.3 &  6.63 & 3.84 & 0.58 $\pm$ 0.11 &  3.3 & 	 16.8\\
 61 & 1 33 44.04 & 30 39 00.0 & -162.4 &  6.48 & 1.72 & 0.27 $\pm$ 0.06 &  2.3 & 	 23.6\\
 62 & 1 34 04.38 & 30 39 22.5 & -177.3 &  3.70 & $<0.20$ & $<0.05$ &  2.2 & 	  3.9\\
 63 & 1 34 10.76 & 30 35 07.5 & -155.5 &  2.68 & $<1.34$ & $<0.50$ &  2.5 & 	  3.5\\
 64 & 1 34 10.22 & 30 46 52.5 & -242.6 &  5.28 & $<0.24$ & $<0.05$ &  1.6 & 	  4.6\\
 65 & 1 34 13.10 & 30 39 07.5 & -197.0 &  4.21 & 0.90 & 0.22 $\pm$ 0.06 &  2.3 & 	  2.7\\
 66 & 1 34 07.31 & 30 47 52.5 & -256.2 & 10.21 & 2.93 & 0.29 $\pm$ 0.05 &  2.8 & 	 11.6\\
 67 & 1 33 52.76 & 30 40 30.0 & -209.0 &  4.05 & 0.36 & 0.09 $\pm$ 0.03 &  2.3 & 	 10.5\\
 68 & 1 34 41.60 & 30 41 22.5 & -208.9 &  6.43 & 0.72 & 0.11 $\pm$ 0.03 &  4.0 & 	  3.3\\
 69 & 1 33 45.78 & 30 40 15.0 & -188.6 &  4.13 & 0.52 & 0.12 $\pm$ 0.04 &  2.1 & 	  5.0\\
 70 & 1 34 14.88 & 30 46 22.5 & -249.6 &  3.03 & 0.40 & 0.13 $\pm$ 0.04 &  2.8 & 	  3.3\\
 71 & 1 33 48.69 & 30 39 45.0 & -174.5 & 10.85 & 3.26 & 0.30 $\pm$ 0.05 &  2.3 & 	 29.0\\
 72 & 1 33 30.12 & 30 32 45.0 & -124.8 &  2.46 & $<0.40$ & $<0.16$ &  1.7 & 	  3.0\\
 73 & 1 33 21.40 & 30 34 00.0 & -131.7 &  5.08 & $<0.70$ & $<0.14$ &  2.4 & 	  0.7\\
 74 & 1 33 45.79 & 30 33 22.5 & -126.2 &  3.82 & 1.76 & 0.46 $\pm$ 0.13 &  1.7 & 	 10.8\\
\end{longtable}

\begin{figure*}[hp]
\begin{center}
\FigureFile(150mm,200mm){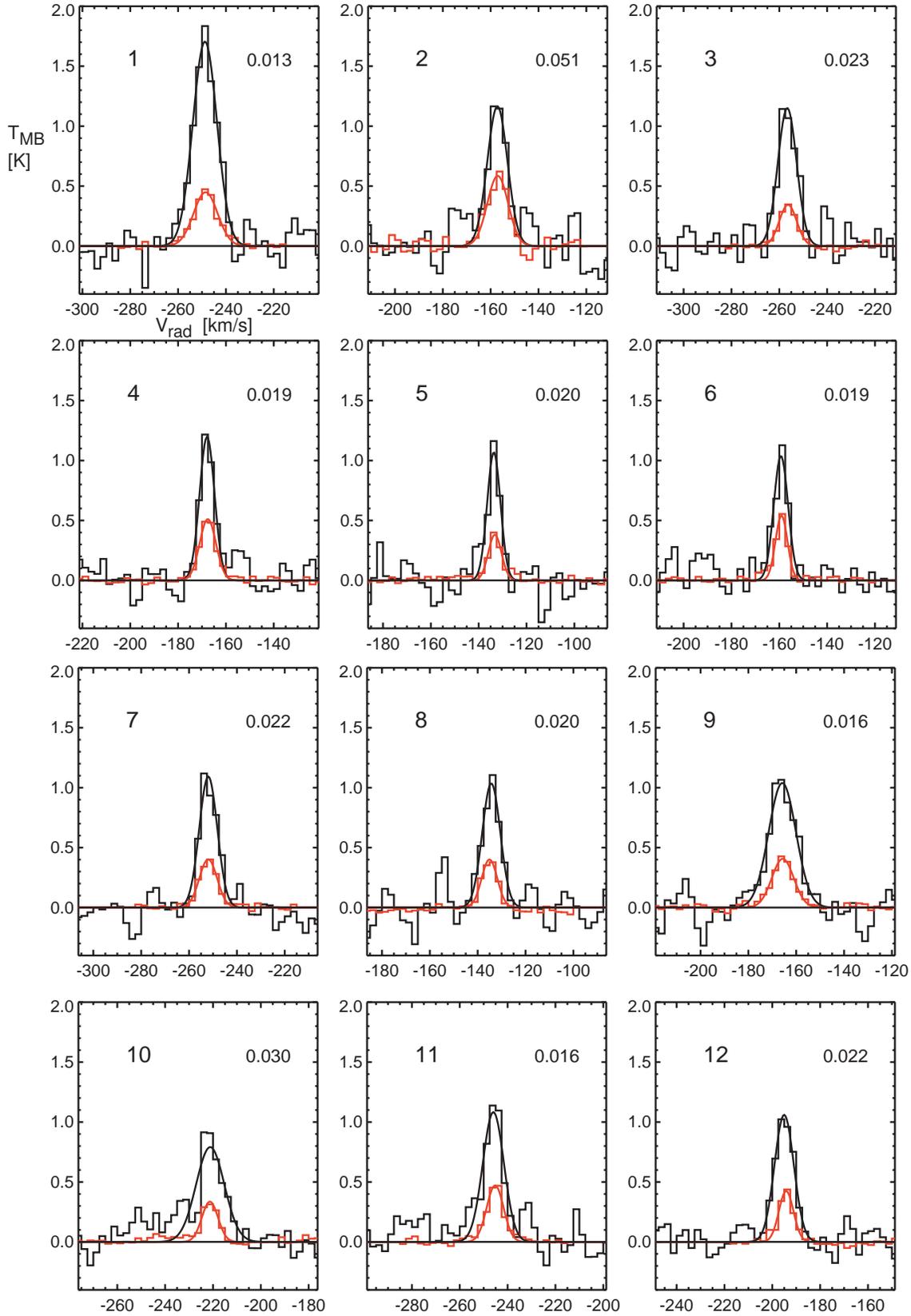}
\end{center}
\caption{$^{12}$CO\,({\it J}\,=\,1--0) (black), and $^{12}$CO\,({\it J}\,=\,3--2) (red) spectra and fitted Gaussians of each GMC on a $T_{\rm MB}$ scale. The $^{12}$CO\,({\it J}\,=\,1--0) images were convolved into a $25''$ ($\sim 100\pc$) beam and regridded to a pixel scale of $7''.5$ in order to match the $^{12}$CO\,({\it J}\,=\,3--2) data. The running number of the GMCs is shown in the upper left and the 1-sigma level of the $^{12}$CO\,({\it J}\,=\,3--2) spectra in a unit of K in the upper right of each panel. Gaussians are not fitted for the $^{12}$CO\,({\it J}\,=\,3--2) spectra in which significant emission is not detected. \label{fig:spectra3210}}  
\end{figure*}
\addtocounter{figure}{-1}

\begin{figure*}[hp]
\begin{center}
\FigureFile(150mm,200mm){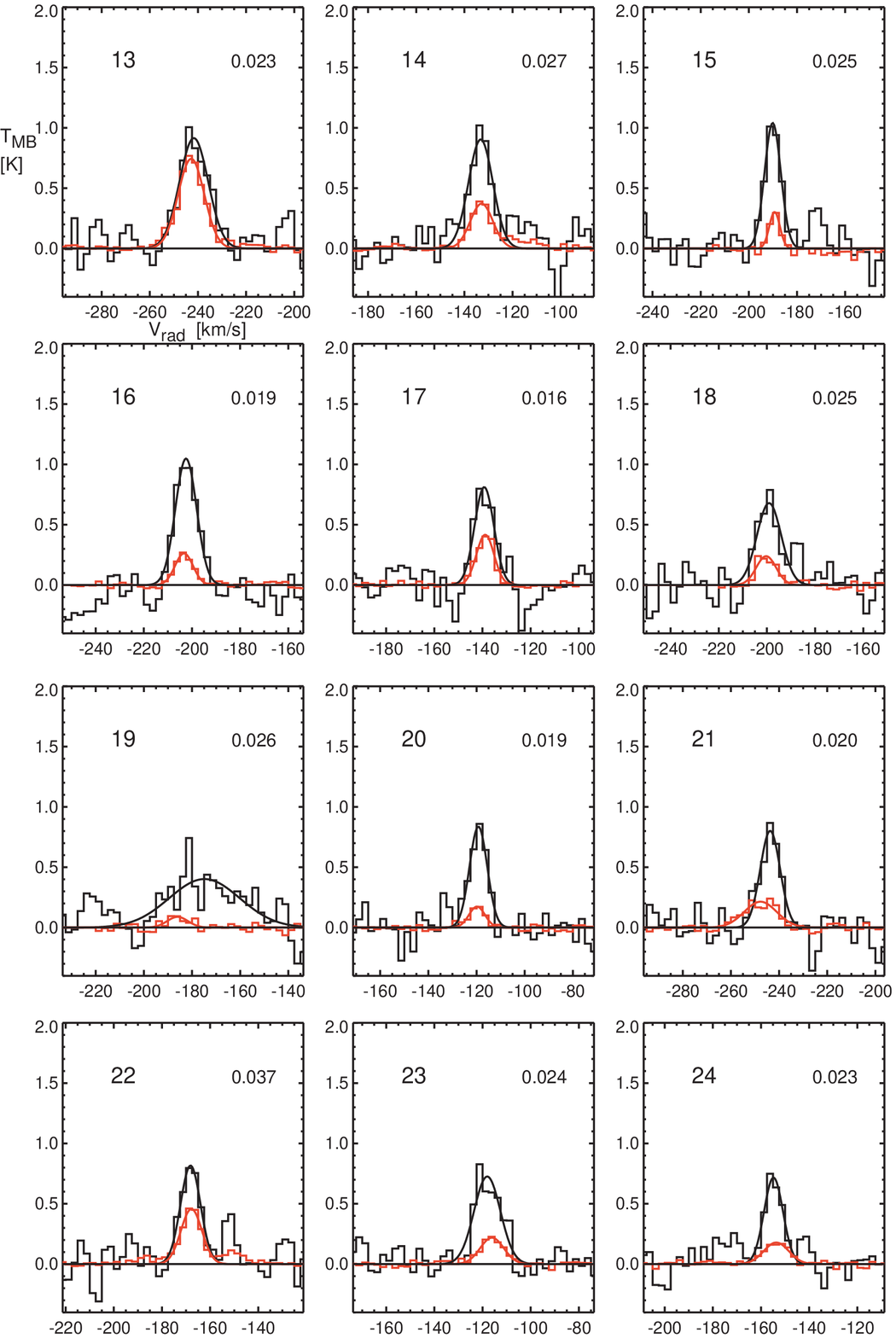}
\end{center}
\caption{(Continued.)}  
\end{figure*}
\addtocounter{figure}{-1}

\begin{figure*}[hp]
\begin{center}
\FigureFile(150mm,200mm){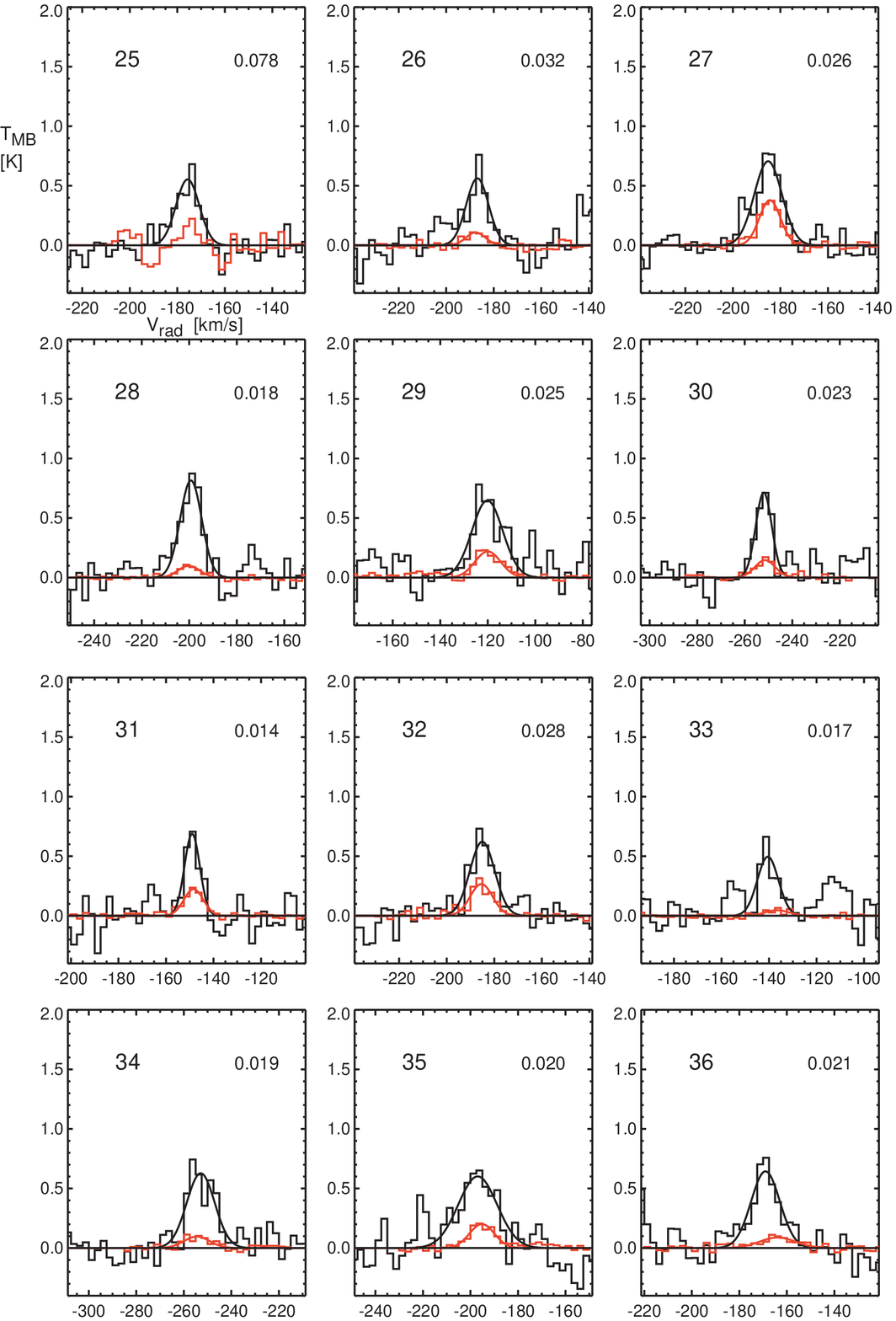}
\end{center}
\caption{(Continued.)}  
\end{figure*}
\addtocounter{figure}{-1}

\begin{figure*}[hp]
\begin{center}
\FigureFile(150mm,200mm){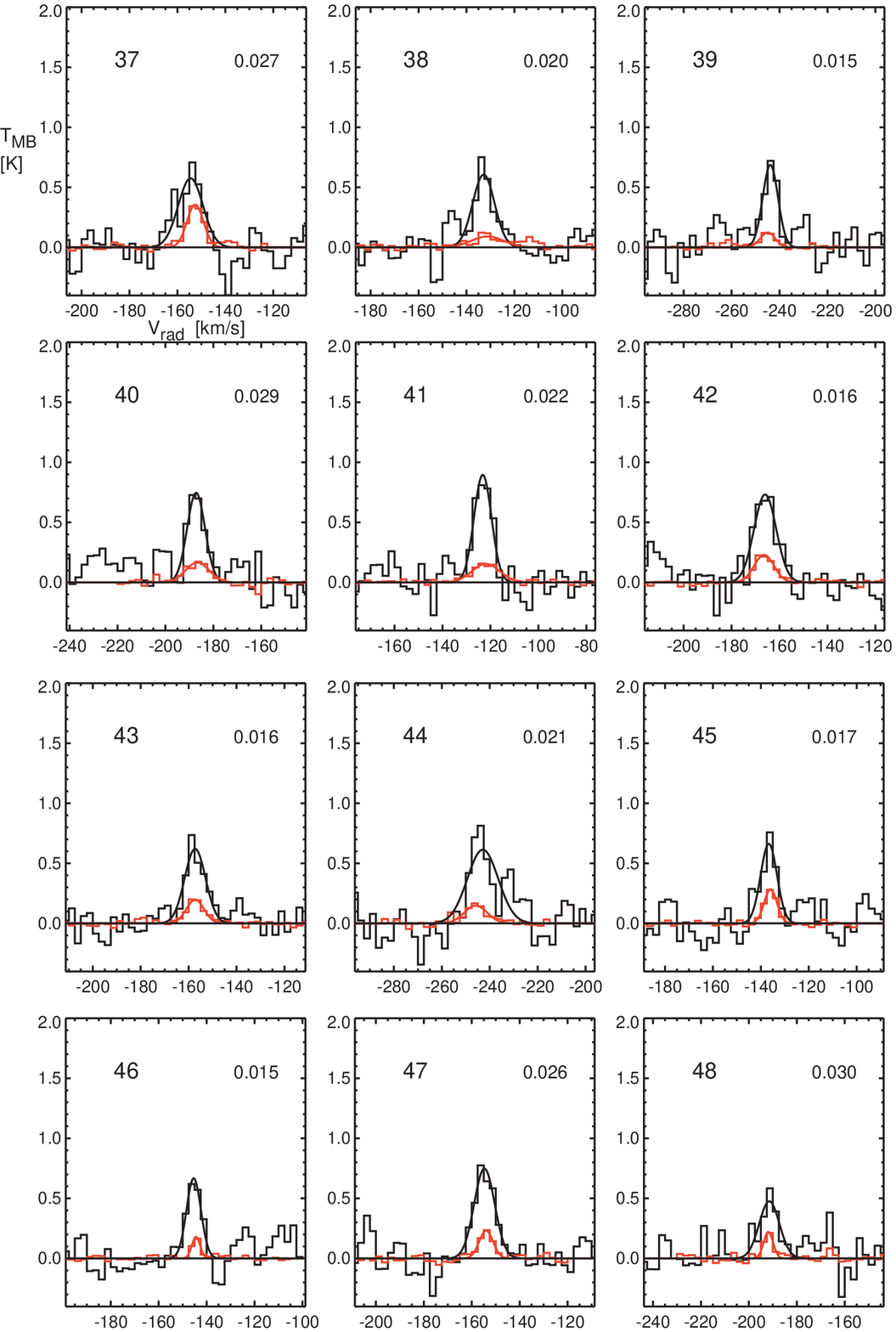}
\end{center}
\caption{(Continued.)}  
\end{figure*}
\addtocounter{figure}{-1}

\begin{figure*}[hp]
\begin{center}
\FigureFile(150mm,200mm){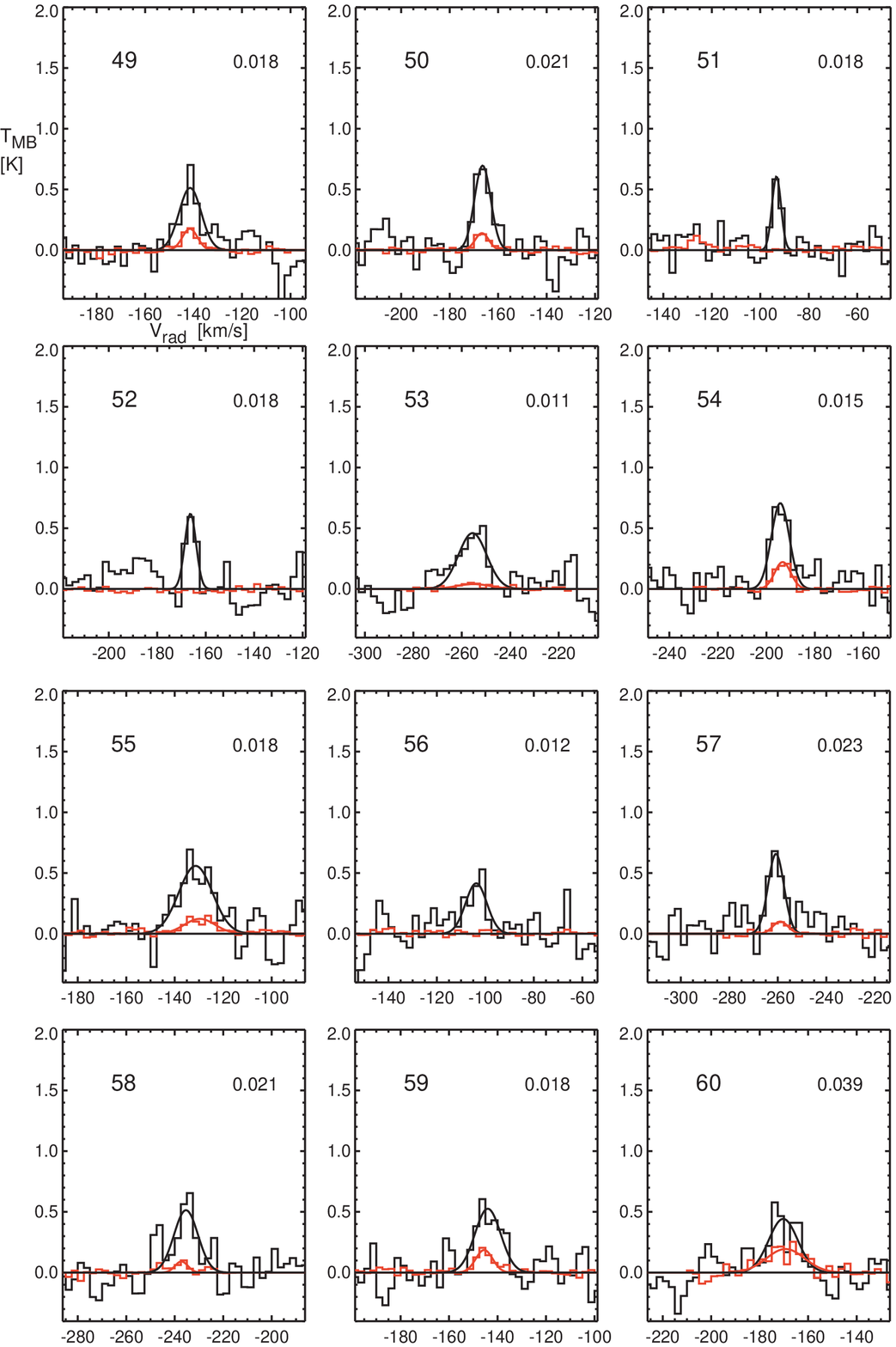}
\end{center}
\caption{(Continued.)}  
\end{figure*}
\addtocounter{figure}{-1}

\begin{figure*}[hp]
\begin{center}
\FigureFile(150mm,200mm){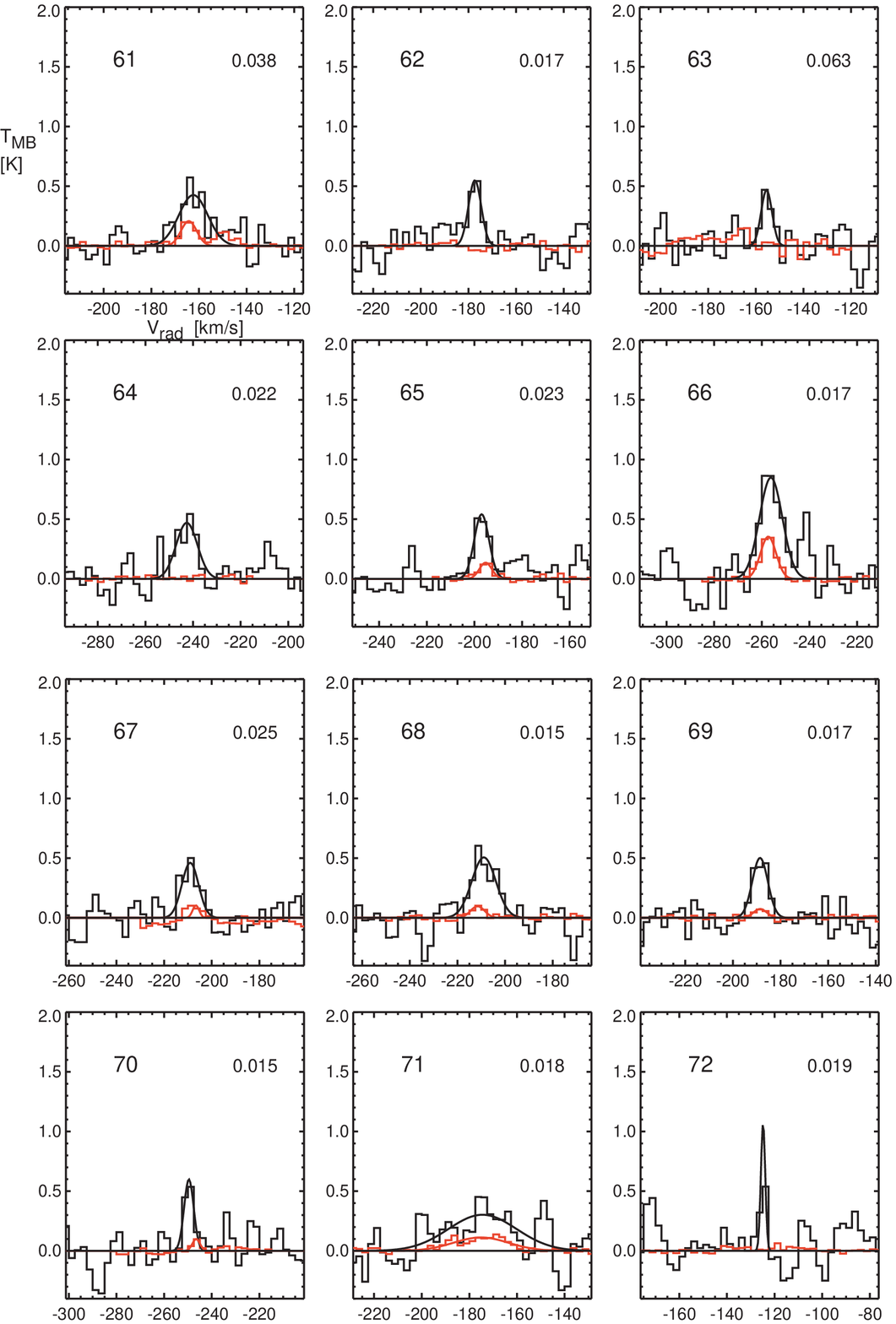}
\end{center}
\caption{(Continued.)}  
\end{figure*}
\addtocounter{figure}{-1}

\twocolumn
\begin{figure*}[hhhhh]
\begin{center}
\FigureFile(100mm,100mm){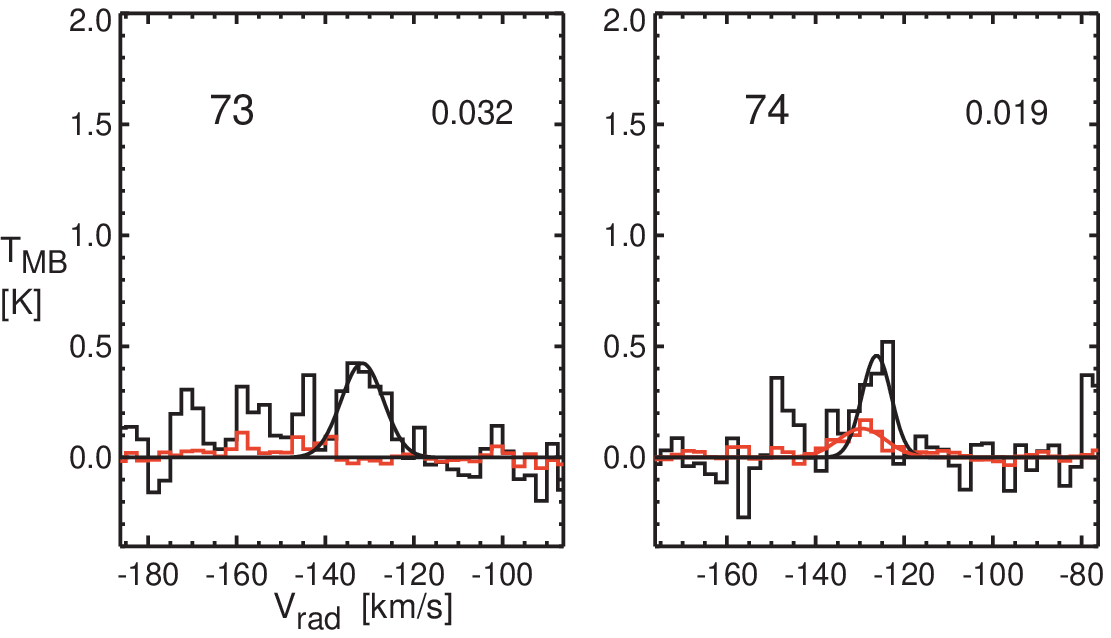}
\end{center}
\caption{(Continued.)}  
\end{figure*}

\begin{figure*}[h]
\begin{center}
\FigureFile(100mm,100mm){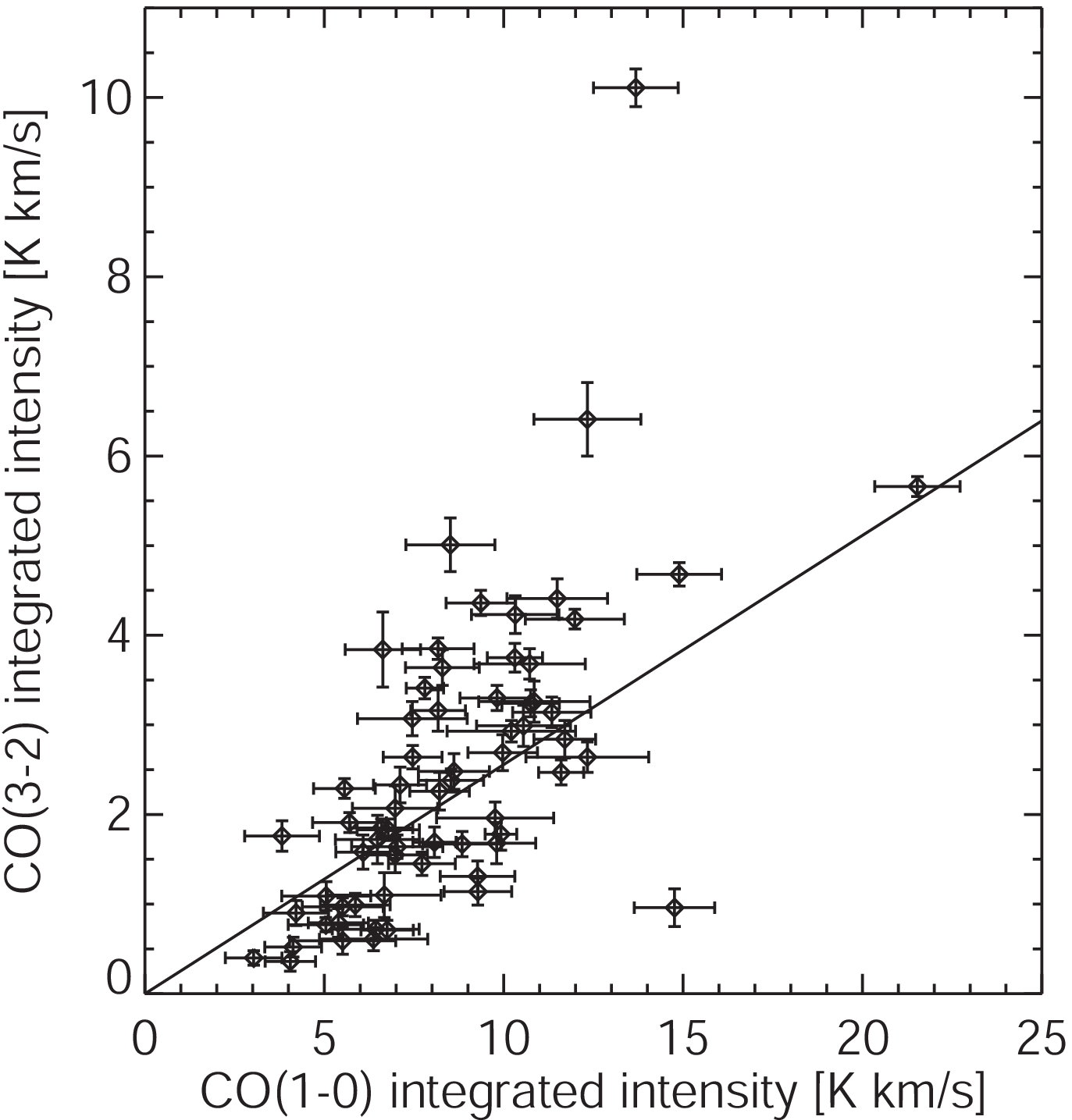}
\end{center}
\caption{Comparison of the $^{12}$CO\,({\it J}\,=\,3--2) and $^{12}$CO\,({\it J}\,=\,1--0) integrated intensities
 (on a $T_{\rm MB}$ scale) at the $T_{\rm CO(1-0)}$ peak position in each GMC. The $^{12}$CO\,({\it J}\,=\,1--0) data were convolved into a $\rm 25''(\sim\! 100\pc)$ beam to match the resolution of the $^{12}$CO\,({\it J}\,=\,3--2) data. The line indicates the weighted-mean value of $R_{3-2/1-0}=0.26$. The GMCs that are not detected for $^{12}$CO\,({\it J}\,=\,3--2) emission are not used. \label{fig:co3210}}  
\end{figure*}

\begin{figure*}[p]
\begin{center}
\FigureFile(140mm,100mm){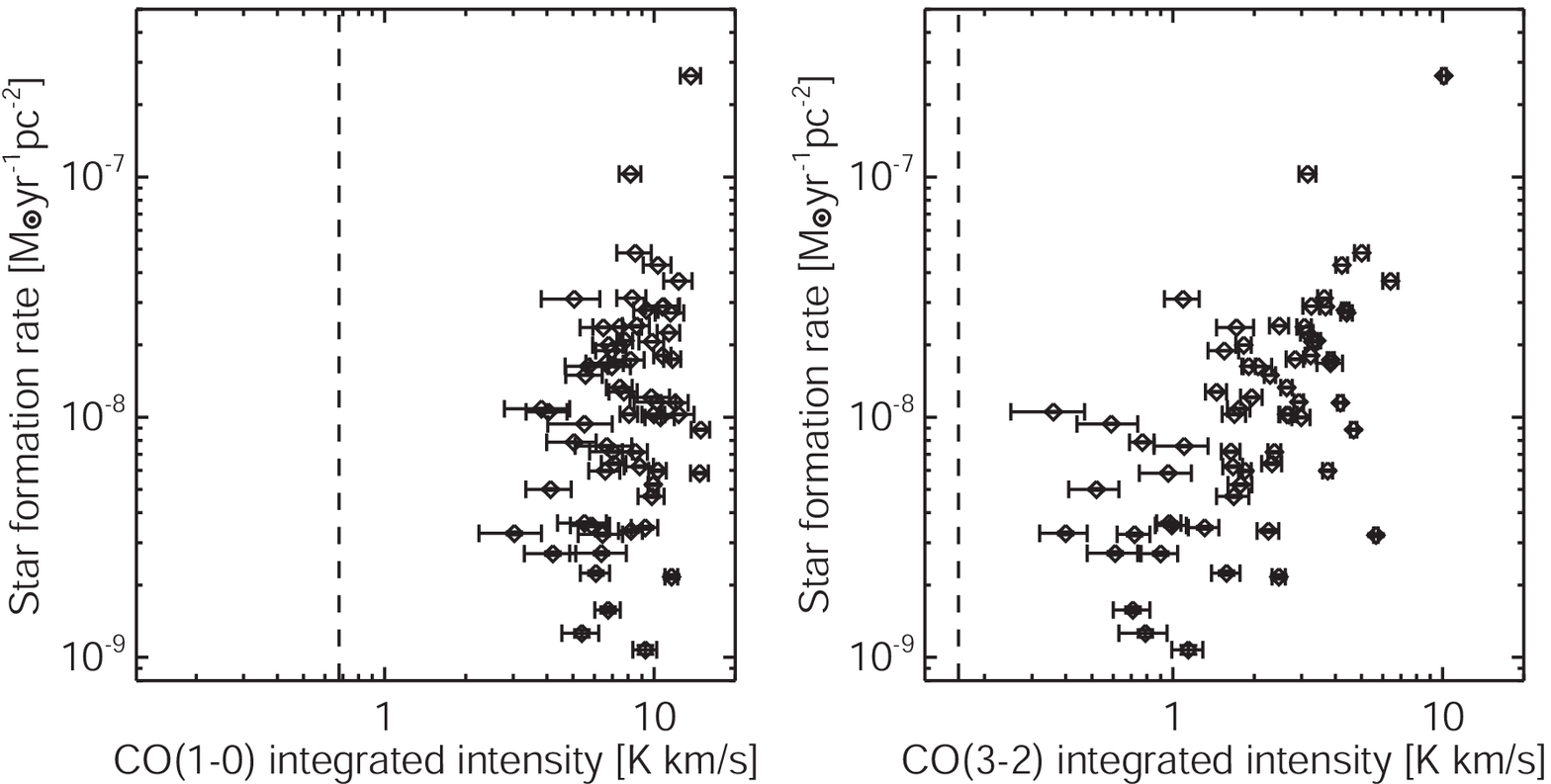}
\end{center}
\caption{The surface density of SFR ($\Sigma_{\rm SFR}$) versus $I_{\rm CO(3-2)}$ and 
$I_{\rm CO(1-0)}$ at the $T_{\rm CO(1-0)}$ peak position 
in each GMC. The $^{12}$CO\,({\it J}\,=\,1--0) and the SFR data were convolved into a
$\rm 25''(\sim\! 100\pc)$ beam to match the resolution of the $^{12}$CO\,({\it J}\,=\,3--2) 
data. The dashed lines indicate the sensitivity limits for each CO line. As mentioned in Section 2, 
the error and the sensitivity limit of $\Sigma_{\rm SFR}$ are 
$4.5\times 10^{-11}{\it \mo}\,\rm \,yr^{-1}\,pc^{-2}$ and  
$1.4\times 10^{-10}{\it \mo}\,\rm \,yr^{-1}\,pc^{-2}$, respectively.
The GMCs that are not detected for $^{12}$CO\,({\it J}\,=\,3--2) emission are not used. \label{fig:COSFR}}   
\end{figure*}

\begin{figure*}[p]
\begin{center}
\FigureFile(100mm,100mm){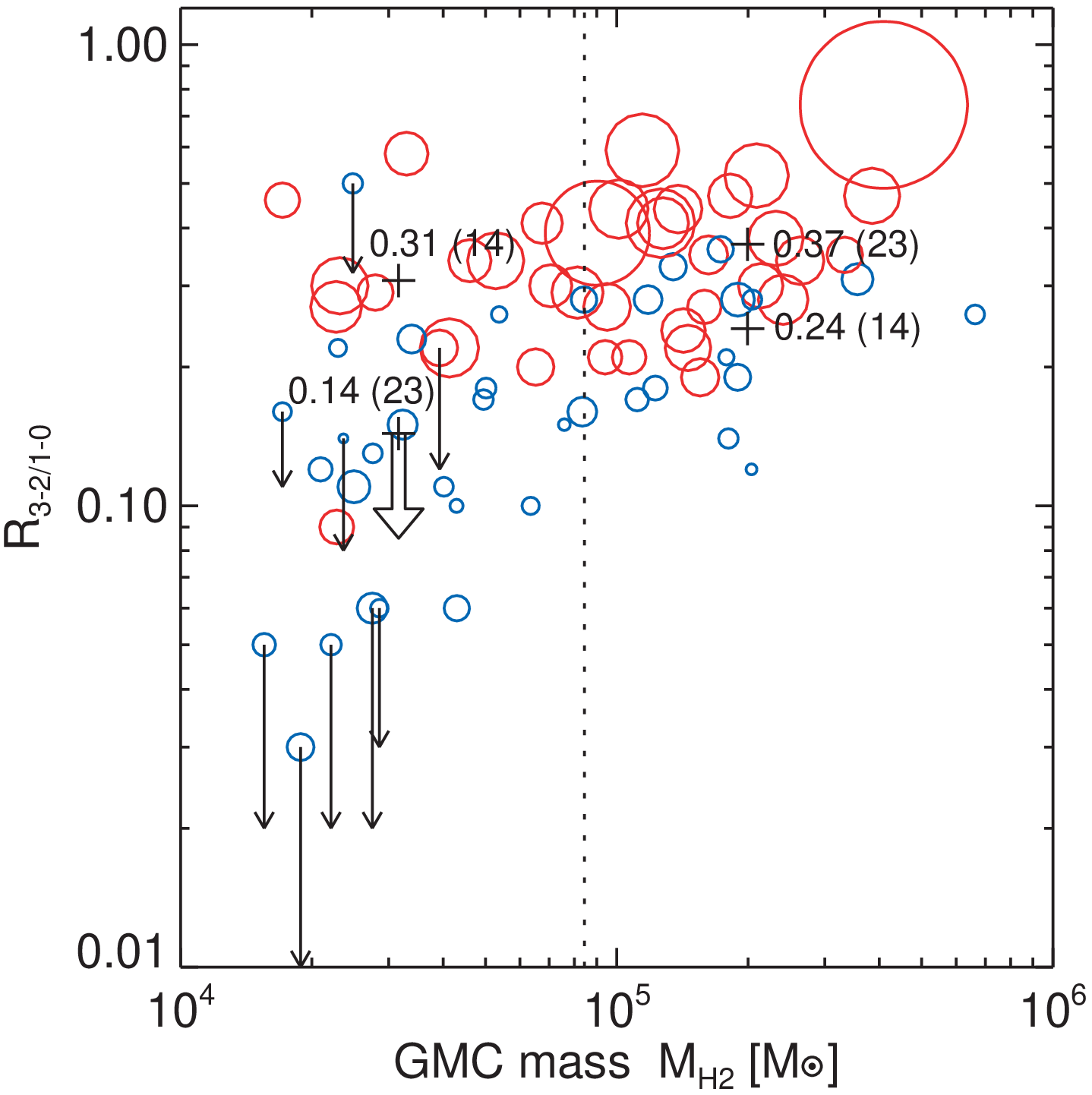}
\end{center}
\caption{GMC mass versus $R_{3-2/1-0}$. The area of the circles is set proportionally to $\Sigma_{\rm
 SFR}$ at the $T_{\rm CO(1-0)}$ peak position of each GMC. The red circles represent GMCs with $\Sigma_{\rm SFR}> 1\times 10^{-8}\mo\pc^{-2}$, 
and the blue ones indicate those with
 $\Sigma_{\rm SFR}< 1\times 10^{-8}\mo\pc^{-2}$. 
The dotted line indicates the criterion of $M_{\rm H2}=8.4\times 10^4\mo$. 
The black crosses and the numbers next to them stand for the averaged $R_{3-2/1-0}$ values 
in each mass-bin, which are taken for GMCs with higher and lower star-forming activity, respectively. 
The numbers in the bracket are the numbers of the GMCs used for averaging.
The $R_{3-2/1-0}$ mean for the less star-forming, less-massive GMCs is an upper-limit value.
The $^{12}$CO\,({\it J}\,=\,1--0) and the SFR data were convolved into a $\rm 25''(\sim\! 100\pc)$ beam to match the resolution of the $^{12}$CO\,({\it J}\,=\,3--2) data. For GMCs whose $^{12}$CO\,({\it J}\,=\,3--2) emissions are not detected, $R_{3-2/1-0}$ is shown as the upper limit, with an arrow whose length is the rms value of $R_{3-2/1-0}$.
 \label{fig:R3210-Mgas}}
\end{figure*}

\end{document}